\documentstyle[12pt]{article}

\renewcommand{\d}{\delta}

\renewcommand{\l}{\lambda}
\newcommand{\m}{\mu}
\newcommand{\n}{\nu}

\renewcommand{\o}{\omega}

\newcommand{\r}{\rho}

\newcommand{\G}{\Gamma}

\newcommand{\mn}{{\mu\nu}}
\newcommand{\mnr}{{\mu\nu\rho}}

\newcommand{\real}{{{\rm I} \kern -.19em {\rm R}}}
\newcommand{\tr}{{\rm {Tr} \,}}

\newcommand{\pa}{\partial}

\newcommand{\intd}{{\int d^4 \! x \,\,}}

\newcommand{\hq}{{{\bf \hat q}}}

\newcommand{\disp}{\displaystyle}
\newcommand{\pad}[2]{{\disp{\frac{\partial #1}{\partial #2}}}}
\newcommand{\fud}[2]{{\disp{\frac{\delta #1}{\delta #2}}}}

\newcommand{\ie}{{{\em i.e.},\ }}

\newcommand{\Lp}{\displaystyle{\biggl(}}
\newcommand{\Rp}{\displaystyle{\biggr)}}

\newcommand{\be}{\begin{equation}}
\newcommand{\ee}{\end{equation}}
\newcommand{\een}[1]{\label{#1}\end{equation}}
\newcommand{\beq}{\begin{eqnarray}}
\newcommand{\eeq}{\end{eqnarray}}
\newcommand{\eeqn}[1]{\label{#1}\end{eqnarray}}
\newcommand{\ba}{\begin{array}}
\newcommand{\ea}{\end{array}}
\newcommand{\brcl}{\begin{equation}\begin{array}{rcl}}
\newcommand{\ercl}{\end{array}\end{equation}}
\newcommand{\ercln}[1]{\end{array}\label{#1}\end{equation}}

\newcommand{\equ}[1]{(\ref{#1})}

\makeatletter
\@addtoreset{equation}{section}
\makeatother

\begin{document}
\setlength{\baselineskip}{3.2ex}
\null\noindent{\normalsize December 1995} \hfill {\normalsize NEIP-95-015}\\
\null\noindent \hfill {\normalsize hep-th/9604020}\\
\begin{center}
{\bf HARD THERMAL LOOPS, QUARK-GLUON PLASMA RESPONSE\\
AND $T=0$ TOPOLOGY\footnotemark[1]}
\footnotetext[1]{ \setlength{\baselineskip}{0pt} Talk given at THERMO95,
the 4th International Workshop on Thermal Field Theories and
Their Applications, Dalian (P.R. China), August 6-10, 1995.
This work has been supported in part by the Swiss National
Science Foundation.}

\vspace*{6.0ex}

{Claudio LUCCHESI}

\vspace*{1.5ex}

{\it Institut de Physique, Universit\'e de Neuch\^atel}\\
{\it 1 rue Breguet, CH -- 2000 Neuch\^atel (Switzerland)}\\
{E-mail: lucchesi@iph.unine.ch}

\vspace*{6.5ex}

{\bf Abstract}

\vspace*{2.0ex}
\parbox{12.8cm}{
\small
I outline various derivations of the non-Abelian Kubo equation, which governs
the response of a quark-gluon plasma to hard thermal perturbations. In the
static case, it is proven that gauge theories do not support hard thermal
solitons. Explicit solutions are constructed within an $SU(2)$ Ansatz and they
are shown to support the general result. The time-dependent problem, \ie
non-Abelian plasma waves, has not been completely solved.  We express and
motivate the hope that the intimate relations linking the gauge-invariance
condition for hard thermal loops to the equation of motion for $T=0$,
topological Chern-Simons theory may yield new insight into this field.
}
\end{center}
\vspace*{1.5ex}

\noindent{\large\bf 1. \hspace*{2.6mm}The Quark-Gluon Plasma}
\setcounter{equation}{0}\setcounter{section}{1}
\vspace{1mm}

The study of the quark-gluon plasma (QGP) phase of high-temperature ${\rm
QCD}_4$ is relevant both to to big-bang cosmology and to heavy ions collisions.
Experiments at CERN and Brookhaven have provided compelling evidence for the
existence of an intermediate ``phase" of QCD during such collisions, with an
energy density $\sim\ 1\ GeV/fm^3$ (see for instance \cite{QGP}). It is however
not clear if this ``phase" is a thermodynamically distinct one.

A heavy-ions interaction, after the primary collision of ions bunches, forms a
thermalized state which lasts for most of the total interaction time. It is in
this thermalized state that the QGP ``phase" is thought to be formed. The
interaction ends by hadronization of the thermalized constituents. Each step of
the process can be tested specifically \cite{QGP,QGPprobes}. However, the
methods available for analysis are empirical, and one feels compelled to work
towards developing a more satisfactory theoretical framework for the
investigation of the QGP.

\vspace{2mm}\noindent{\large\bf 2. \hspace*{2.6mm}High-Temperature Perturbative
QCD}
\setcounter{equation}{0}\setcounter{section}{2}
\vspace{1mm}

High-temperature perturbative QCD provides a natural setting for the
theoretical study of the QGP. We shall adopt the hypotheses that (i) QCD is in
its deconfined phase and that (ii) the plasma is collisionless, so that it can
be regarded as a dilute gas of quarks and gluons. The dominant interaction is
between hard colored plasma particles and the soft mean color field. ``Hard"
denotes an energy scale of the order of the temperature, $\varepsilon\sim T$,
while ``soft" refers to the energy scale $\varepsilon\sim gT$, $g$ being the
gauge coupling constant.
It is to expect that motion on a distance scale $\sim 1/gT$, the inverse soft
energy scale, should involve coherently many hard particles, giving rise to
color polarization effects in the QGP. Indeed, the real part of the color
polarization tensor describes Debye screening of the color charges, while its
imaginary part accounts for Landau damping of the mean color field.

If zero-temperature perturbative QCD is naively extended to the $T > 0$ case,
physical quantities turn out to be gauge-dependent. This old puzzle has been
solved by the discovery of the so-called ``hard thermal loops" (HTLs)
phenomenon \cite{ref3}: at high temperature, effects of leading order in $g$
can arise at any order of perturbation theory, \ie loop diagrams exist that are
comparable to the corresponding tree amplitude. Such diagrams contain one-loop
integrals with soft external lines and hard internal (loop) momenta, and are
termed ``hard thermal loops". In order to account consistently for the orders
in $g$, one is led to resum the perturbative expansion.

Hard thermal loops represent thermal, not quantum, corrections. They describe
polarization effects in the plasma, as well as the propagation of plasma waves,
plasma solitons, etc.
The hard thermal loops are the dominant diagrams at high temperature. They are
proportional to the Debye mass squared
$m_D^2= {g^2\,T^2\over 3}\ (N+{N_F\over 2})$, where $N$ is the number of colors
and $N_F$ the number of flavors. Note that the Debye mass is proportional to
the soft scale $gT$; $m_D$ represents the smallest possible frequency for
time-dependent (gluon) propagation. The Debye screening length is just its
inverse, $\l_D=m_D^{-1}\sim 1/g\,T$.

\vspace{2mm}\noindent{\large\bf 3. \hspace*{2.6mm}The Non-Abelian Kubo
Equation}
\setcounter{equation}{0}\setcounter{section}{3}
\vspace{1mm}

Hard thermal loops are generated by an effective action \cite{ref3,ref2}:
\be
\G_{\rm HTL} (A) = {m_D^2\over 2}\left[  \intd A_0^a A_0^a + \int
{d\Omega_{\hq} \over 2\pi} \  W(A_+)\right]\ .
\label{genfctl}
\ee
The first term in the right side describes color-electric screening, and
$W(A_+)$ is a non-local functional of $A_+\equiv {1\over\sqrt{2}}(1;\hq)\cdot
A$, where $\hq$ is an arbitrary unit 3-vector, $\hq^2=1$. The measure
$d\Omega_{\hq}/ 2\pi$ denotes integration over the solid angle spanned by
$\hq$.

The response of the QGP to an external perturbation is governed by a Yang-Mills
field equation
\be
D_\m \,F^{\m\n}={m_D^2\over 2}\ J^\m_{\rm induced}\ ,
\label{nake1}
\ee
with the induced hard thermal current
${m_D^2\over 2}\,J^\m_{\rm induced}\equiv -{\d\over \d A_\m} \G_{\rm HTL} (A)$
as its source\footnote{Note that \equ{nake1} is just the equation of motion for
the gauge field, when one adds the zero-temperature QCD action to the HTLs
generating functional, \ie \equ{nake1} is equivalent to requiring
${\d\over \d
A_\m}( S_{\rm QCD}^{\,T=0}+\G_{\rm HTL}) =0$.}.
In order to close the system of differential equations for $A_\m$, we need a
constraint on the current. Exploiting the analogy between the gauge invariance
of HTLs and Chern-Simons theory\footnote{See Section 5 below for a more
detailed account of this analogy.} \cite{ref1}, one derives the induced current
in the form \cite{ref5}
\be
J_{\rm induced}^\mu = \int {d \Omega_\hq \over 4\pi}
\, \biggl( Q^\mu_{+} \biggl[ a_{-}(x) - A_{-}(x) \biggr] +
Q^\mu_{-} \biggl[ a_{+}(x) - A_{+}(x) \biggr] \biggr)\ ,
\label{nake2}
\ee
where $Q^\mu_{\pm}\equiv {1\over\sqrt{2}} (1, \pm \hq)$,  $\hq^2=1$, are
light-like 4-vectors,
$A_{\pm}= Q^\mu_{\pm} \, A_\mu$ are light-like projections of $A_\mu$, and
$a_\pm$ are constrained, introducing $\pa_\pm=Q^\m_\pm\pa_\m$, by
\be
\partial_+ a_- - \partial_- A_+ + [A_+, a_-] = 0\ ,\qquad
\partial_+ A_- - \partial_- a_+ - [A_-, a_+] = 0\ . \label{nake3}
\ee
{}From now on, we shall call {\it non-Abelian Kubo equation} the system
\equ{nake1}--\equ{nake3}, which governs hard thermal processes in the QGP.

Alternative derivations of the non-Abelian Kubo equations have been given.

{\bf 1.} The non-Abelian Kubo equation can be derived from quantum kinetic
theory \cite{ref8}. In that approach, one starts from the Schwinger-Dyson
equations, and uses a kinematic approximation scheme in which derivatives with
respect to center-of-mass and to relative coordinates carry different
$g$-dependences. From the resulting quantum kinetic equations, one is able to
derive $\Gamma_{\rm HTL}(A)$, and hence an expression for the induced current.

{\bf 2.} The non-Abelian Kubo equation can also be derived, as done in
\cite{JLL}, by applying the action principle to the Cornwall-Jackiw-Tomboulis
composite effective action $\Gamma[\phi,G_\phi]$ \cite{ref7}, a functional of a
generic field $\phi$ (which stands for gluons, quarks, antiquarks, ghosts, etc)
and of the two-point functions $G_\phi$. Requiring that $\Gamma[\phi,G_\phi]$
be stationary with respect to $\phi(x)$ and $G_\phi(x,y)$ yields, respectively,
the equation \equ{nake1} and a set of conditions on the two-point functions
which can be cast in the form of a constraint on the induced current
\cite{JLL}.

This approach is close to the one of \cite{ref8}, of which it adopts numerous
steps, including its kinematic approximation scheme. The composite action
approach presents the advantage of providing an action principle for hard
thermal processes in the quark-gluon plasma. Whether, and which, physical
hypotheses can be expressed at the level of the composite action functional is
an open question.

{\bf 3.} More recently, the non-Abelian Kubo equation has been derived from
classical transport theory \cite{KLLM}. This is the subject of the talk by C.
Manuel at this meeting. One starts from classical particle dynamics as
described by the Wong equations; the color charge is considered as a {\it
classical} phase-space variable. Transport theory yields the non-Abelian Vlasov
equations, and the induced current is expressed in terms of the one-particle
distribution functions. Expanding these distribution functions in powers of the
coupling constant $g$ yields, at leading order, the gauge invariance condition
for the generating functional of hard thermal loops $\Gamma_{\rm HTL}(A)$
\equ{genfctl}. Subsequent developments that lead to the form
\equ{nake2}--\equ{nake3} for the induced current, and hence to the non-Abelian
Kubo equation, are described in \cite{ref5}.

\vspace{2mm}\noindent{\large\bf 4. \hspace*{2.6mm}Solutions of the Non-Abelian
Kubo Equation: QGP \\ \hspace*{1cm}Response}
\setcounter{equation}{0}\setcounter{section}{4}
\vspace{1mm}

It is of obvious interest to discuss solutions of the non-Abelian Kubo equation
\equ{nake1}--\equ{nake3}. In the
Abelian, electrodynamical case this is easy to do, since the conditions
\equ{nake3} can be readily solved for $a_\pm$, and
the solutions of the linear problem are the well-known Abelian plasma waves
\cite{ref6}. The non-linear problem of finding non-Abelian plasma waves
 is much more formidable. Also, one inquires whether the non-linear equations
support soliton (and instanton) solutions.

Let us first consider the static case.
Acting on time-independent fields $A_\pm=A_\pm({\bf q})$, the derivatives in
\equ{nake3} become $\partial_\pm =\pm {1\over\sqrt{2}} \hq \cdot {\bf \nabla}
\equiv \pm\partial_{\bf q}$; therefore the constraints \equ{nake3} can be
written as
\be
\partial_{\bf q} \, (A_+ + a_-) + [A_+, A_+ + a_-] = 0\ ,\qquad
\partial_{\bf q} \, (A_- + a_+) - [A_-, A_- + a_+] = 0\ .
\label{2.2}
\ee
These constraints are solved by $a_- = -A_+$, resp. $a_+ = -A_-$. \equ{nake2}
then yields
\be
J_{\rm induced}^\mu({\bf q})= - \int {d\Omega_\hq \over 4\pi} \,
( Q^\mu_{+} + Q^\mu_{-} ) \,( Q^\nu_+ + Q^\nu_- ) A_\nu({\bf q})
\label{2.4}
\ee
for the static inducent current. With ${\bf Q}_{+} + {\bf Q}_- = 0$ and
$Q_{+}^0 + Q_{-}^0 = \sqrt{2}$, one computes $J_{\rm induced}^\mu = (- 2\,
A^0;{\bf 0})$. The response equations \equ{nake1} become, in the static limit:
\be
{D}_i E^i= - m_D^2 A_0  \ ,\qquad
\epsilon^{ijk} {D}_j B^k -[A_0, E^i]= 0\ , \label{2.6}
\ee
where $E^i \equiv F^{i 0}$ and $F^{ij} \equiv -\epsilon^{ijk} B^k$, and the
Debye mass $m_D$ plays the role of a gauge invariant, color-electric screening
mass.

In the study of static excitations, one is of course interested in the issue of
possible (static) soliton solutions. It has been established in \cite{JLL} that
the non-Abelian Kubo equation does not possess any finite energy static
solutions. The proof goes as follows. Consider the symmetric tensor
\be
\theta^{ij} = 2\, \tr \biggl( E^i E^j + B^i B^j - {\delta^{ij} \over 2}
(E^2 + B^2 + m_D^2 A_0^2) \biggr)\ .
\label{2.7}
\ee
Using (\ref{2.6}) one verifies that for static fields
$\partial_j \, \theta^{ji} = 0$.  Therefore
\be
\int d^3 r \, \theta^{i i} =  \int d^3 r \, \partial_j (x^i \, \theta^{ji})=
\int dS^j x^i \theta^{ji}\ .
\label{bull}
\ee
Moreover, the energy of a massive gauge field (with no mass for the
spatial components) can be written as
\be
{\cal E} = \int d^3 r \biggl\{ - \tr
\biggl( E^2 + B^2  + {1 \over m_D^2} (D_i E^i)^2 \biggr)
+ \tr \biggl( m_D A_0 + {D_i E^i \over m_D} \biggr)^2 \biggr\}\ .
\label{2.9a}
\ee
The second trace in the integrand enforces the first of constraints
(\ref{2.6}). Consequently, on the constrained surface the energy is a sum of
positive terms,
\be
{\cal E} = \int d^3 r
           \left\{ - \tr \left( E^2 + B^2 + m_D^2 A_0^2 \right) \right\}\ ,
\label{2.9b}
\ee
and ${\bf E}$, ${\bf B}$ and $A_0$ must decrease at large distances
sufficiently rapidly so that each of them is square integrable. This in turn
ensures that the surface integral at infinity in (\ref{bull}) vanishes, so that
static solutions require
$\int d^3 r \ \theta^{i i} = 0$.
On the other hand, from (\ref{2.7}), we see that $\theta^{ii}$ is a sum
of positive terms, $\theta^{ii} = -\tr ( E^2 + B^2 + 3 m_D^2 A_0^2 )$,
hence we must have $\bf E= B =0$ and $A_0=0$, \ie there are no (non-trivial)
finite-energy solutions to \equ{2.6}.
A similar argument \cite{JLL} shows the absence of ``static" instantons.

I want to analyze now in some details possible solutions of the static response
equations, and present numerical results \cite{JLL} that support the general
proof of absence of (static) solitons. I consider for simplicity the radially
symmetric restriction of the static response equations \equ{2.6}, in the
$SU(2)$ case. Radially symmetric $SU(2)$ gauge potentials have the
form\footnote{A residual gauge freedom has been used to eliminate
a term proportional to $ \hat{r}^a \hat{r}^i$.}
\be
\label{eq:a1}
A_0^a = \hat{r}^a \, { g(r) \over r}\ , \qquad
A^a_i = ( \delta^{ai} - \hat{r}^a \hat{r}^i ) \, {\phi_2(r) \over r} +
\varepsilon^{aij} \, \hat{r}^j \, {{1 - \phi_1(r)} \over r}\ .
\ee
We substitute this {\it Ansatz} into (\ref{2.6}).
The resulting equations
give us the freedom to set $\phi_2$ to zero. Rescaling $x=mr$ and defining
$J(x)=g(r)$, $K(x)=\phi_1(r)$, we obtain a system of coupled second-order
differential equations
\be
x^2 {d^2 \over dx^2}\, J = (x^2 + 2 K^2) \, J\ , \qquad
x^2 {d^2 \over dx^2}\, K = (K^2 -J^2 -1) \, K\ .\label{a3}
\ee
These equations possess two exact solutions:\hfill\break
(1) the {\it Yang-Mills vacuum}, with $J=0$, $K=\pm 1$, and \hfill\break
(2) the {\it Wu-Yang monopole}, plus a screened electric field:
$J=J_0\,e^{-x}$, $K=0$.

Let us investigate the asymptotic behaviour of the system \equ{a3}. At
$x\rightarrow\infty$, the
regular solution tends to the Yang-Mills vacuum, with $J$ approaching its
asymptote exponentially\footnote{There is also a solution with $J$ growing
exponentially, which we do not consider.}.
Near the origin, \ie at $x\rightarrow 0$, $J$ and $K$ behave either like the
Yang-Mills vacuum, or
approach the Wu-Yang monopole, as
\be
J(x) \rightarrow J_0 + ... \ ,\qquad
K(x) \rightarrow K_0\sqrt{x} \, {\rm cos}\biggl( {\sqrt{J_0^2+{3\over 4}}}
\ ln {x \over x_0} \biggr)+ ...\ .
\label{eq:a44}
\ee

Only the vacuum alternative at the origin could lead to finite energy.
However, since we must choose one of two possible solutions at infinity
(obviously we pick the regular one), the behavior at
the origin is determined and can be exhibited explicitly by integrating the
equations (\ref{a3}) numerically. Starting
with regular boundary conditions at infinity, we find that the monopole
solution is reached at the origin, with $K$ vanishing as in (\ref{eq:a44}); see
Figure 1 below.
This result is consistent with the analytic proof \cite{JLL} that there are no
fi\-nite energy static solutions in hard thermal gauge theories.

Much less is known about time-dependent solutions of the non-Abelian Kubo
equation \equ{nake1}--\equ{nake3}. The special case of time-dependent,
space-independent solutions has been investigated in \cite{BInew}. The {\it
Ansatz} $A^\m(x)=(0,{\bf A}(t))$ leads to the induced current $J^\m_{\rm
induced} = ( {0;-{2\over 3}\,{\bf A}})$, which describes global color
oscillations in the plasma.
A more general {\it Ansatz}, $A_\m(x)=A_\m({\bf p}\cdot{\bf x}-\omega t)$, has
been presented in \cite{BInew2} to describe non-Abelian plane waves in the
plasma.

\vspace*{-2cm}
\hspace{-1.3cm}
\let\picnaturalsize=N
\def\picsize{5.5in}
\def\picfilename{dalianfigure.ps}
\ifx\nopictures Y\else{\ifx\epsfloaded Y\else\input epsf \fi
\let\epsfloaded=Y
\centerline{\ifx\picnaturalsize N\epsfxsize \picsize\fi
\epsfbox{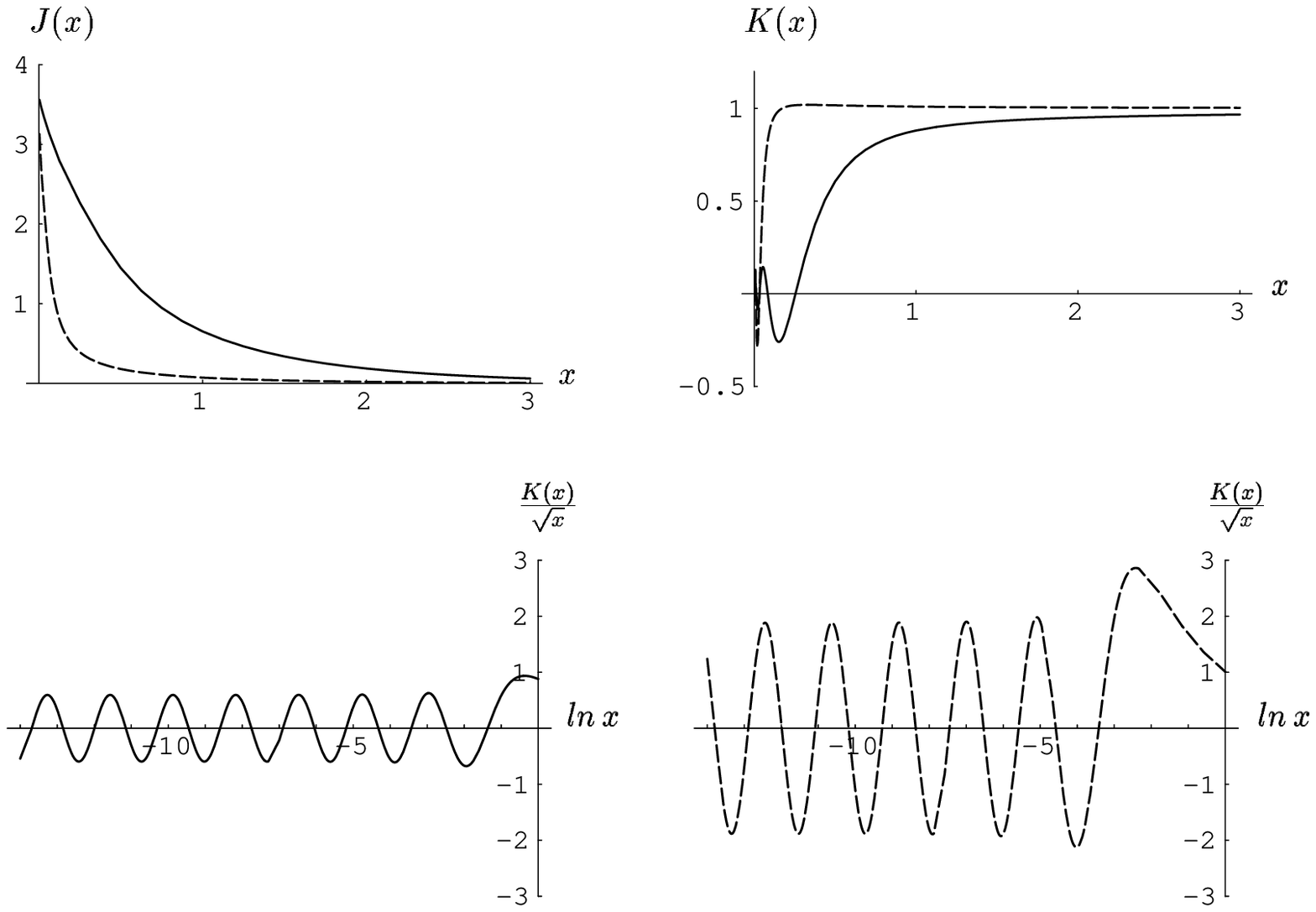}}}\fi
\vspace*{-7.5cm}
{\bf Figure 1}:  $J(x)$ and $K(x)$ obtained by integrating numerically
equations \equ{a3}, starting with regular boundary conditions at infinity. The
plain and dashed lines represent different rates of approach to the (Yang-Mills
vacuum) asymptotes at $x=\infty$, with $K=1$. Analysis of the oscillations in
$K$ near the origin [see the plots of $K(x)/\sqrt{x}$ vs. ${\rm ln}(x)$]
verifies the analytical form of the aymptotes \equ{eq:a44}.

\vspace{2mm}\noindent{\large\bf 5. \hspace*{2.6mm}QGP Response and $T=0$
Topology}
\setcounter{equation}{0}\setcounter{section}{5}
\vspace{1mm}

In order to solve completely the time-depependent non-Abelian Kubo equation,
one may need new insight, a possible source of which is provided by
zero-temperature, topological gauge theory. Indeed, the gauge invariance
condition for the generating functional of hard thermal loops in ${\rm QCD}_4$
can be identified with the equation of motion for the gauge field in $T=0$,
$d=3$ topological Chern-Simons theory \cite{ref1}.

Let us see how this works in some details. We first perform a gauge
transformation $\d_g A_\m =\pa_\m\omega + [A_\m,\omega]$, with parameter
$\omega=-it^a\omega^a$, of the HTLs generating functional \equ{genfctl}.
Imposing gauge invariance of \equ{genfctl}, $\d_g \G_{\rm HTL}(A) =0$, implies
\be
\int d\Omega_\hq\ \d_g W(A_+)
= 4\pi\ \intd (\pa_0 A_0^a)\omega^a\ .
\ee
Expressing $\d_g W(A_+)$, upon integrating by parts, as
$$
\d_g W(A_+) =-\intd  \fud{W(A_+)}{A_+}\,\d_g A_+=
-\intd \left\{ \pa_+{{\d W}\over{\d A_+}}
+\left[ A_+,{{\d W}\over{\d A_+}}\right]\right\}^a\o^a
$$
leads, upon introducing the combination
$f\equiv {{\d W}\over{\d A_+}}+A_+$, to the gauge invariance condition for hard
thermal loops \cite{ref2}, written now as:
\be
\pad{f}{u}+[A_+,f]+\pad{A_+}{v}=0\ ,
\label{tralala}
\ee
in the new space-time coordinates $(u,v,{\bf x}^T)$ defined by $u\equiv
Q^\m_-x_\m$, $v\equiv Q^\m_+x_\m$, and ${\bf Q}_+\cdot\, {\bf x}^T=0$.

To establish the relation with a topological action, one proceeds following
\cite{ref1}. First, perform a Wick rotation into Euclidean $\real^4$
space-time, and rename the coordinates $u=z$, $v=\bar z$. Next, introduce the
notation $A_+=a_z$ and define $a_{\bar z}\equiv -f$. Last, enforce the axial
gauge condition $a_0=0$. The gauge invariance condition for HTLs \equ{tralala}
can then be rewritten as
\be
\pa_{\bar z} a_z-\pa_z a_{\bar z} +[a_{\bar z},a_z]=0\ .
\label{zerc}
\ee
This is the zero-curvature condition $F_{z{\bar z}}=0$ for the gauge field
strength tensor $F_{z{\bar z}}$. I now switch to the Landau gauge, $\pa_\m
A^\m=0$, for purposes of illustration. In that gauge, the zero-curvature
condition \equ{zerc} takes the form $F_\mn =\pa_\m A_\n-\pa_\n A_\m
+[A_\m,A_\n]=0$. This is just the equation of motion for the gauge field $A_\m$
with the classical action
\be
S=\int d^3\!x\ \varepsilon^\mnr\Lp A^a_\m\pa_\n A^a_\r +{2\over 3}\,A^a_\m
A^b_\n A^c_\r \,f^{abc}\Rp\ .
\label{csaction}
\ee
The latter action is the one of topological Chern-Simons theory, a
zero-temperature gauge field theory in three space-time dimensions.
``Topological" refers to the fact that the Lorentz indices in \equ{csaction}
are saturated without using the metric tensor $g_{\mn}$. As a consequence, the
energy-momentum tensor, computed as $T^\mn = {\d S\over \d g_\mn}$, doesn't get
any contributions from the action \equ{csaction}. Only the (non-topological,
\ie metric-dependent) gauge fixing and ghost terms that one adds to the action
in order to quantize the theory contribute to $T^\mn$, but such contributions
do not represent physical energy-momentum.

Chern-Simons theory, together with other topological field theories, has been
intensively studied in recent years \cite{BBRT}, both due to its mathematical
and physical interests. Chern-Simons theory describes anyons (particles with
non-integer spins) and is a starting point for modeling the quantum Hall effect
and high-$T_c$ superconductivity. Chern-Simons theory is known to possess the
appealing property of perturbative finiteness \cite{csfiniteness}, which in
turn is connected to an interesting (twisted) form of supersymmetry
\cite{cssusy,csfiniteness} that is peculiar to topological theories. The
question wether such properties could be of interest in the realm of high-$T$
QCD$_4$ is an open, and interesting one. In particular, the twisted topological
supersymmetry of Chern-Simons theory \cite{bdl} could reveal a useful tool to
shed new light on the relationship between high temperature in field theory and
(untwisted) physical supersymmetry.

\vspace{2mm}\noindent{\large\bf 6. \hspace*{2.6mm}Conclusions}
\setcounter{equation}{0}\setcounter{section}{6}
\vspace{1mm}

There are indications that the quark-gluon plasma may have been observed
experimentally in heavy-ions collisions. However, a satisfactory theory of the
QGP phase is still lacking. The best available theoretical framework is
high-temperature QCD, in which hard thermal effects are dominant.

The response of a quark-gluon plasma to a hard thermal perturbations is
described by the non-Abelian Kubo equation. The static case is completely
solved. In particular, there do not exist hard thermal solitons. This general
result is confirmed by numerical computations, within a radial $SU(2)$ {\it
Ansatz}. The time-dependent case is less well understood.

New insight into this field may hence be needed, and could indeed be provided
by the close relationship that exists between hard thermal ${\rm QCD}_4$ and
three-dimensional Chern-Simons theory, a topological gauge theory at zero
temperature. Chern-Simons theory is perturbatively finite. Furthermore, its
gauge-fixed action exhibits a twisted supersymmetry which relates ``bosons"
(fields of even ghost number) to ``fermions" (fields of odd ghost number), in
very close analogy to usual (untwisted) supersymmetry. This leads us to the
following, open questions. (1) Can the twisted supersymmetry of Chern-Simons
theory be translated into the language of high-temperature QCD, and what does
it mean in that context? (2) Can this shed new light onto the interplay between
supersymmetry and high-temperature in field theory?

\vspace{3mm}\noindent{\bf Acknowledgements:}
The author expresses his gratitude to
R. Jackiw and C. Manuel for interesting discussions and
comments on the manuscript.


\end{document}